\documentclass[a4paper,
             ]{paper}

\usepackage{graphicx}

\begin{document}

\title{Reconstruction of longitudinal electrons bunch profiles at FACET, SLAC\thanks{Work supported by funding from Universite Paris Sud, program "Attractivite" and by the ANR under contract ANR-12-JS05-0003-01}}

\author{J.Barros, N.Delerue, S.Jenzer, M. Vieille Grosjean\thanks{vgrosjea@lal.in2p3.fr}, LAL, Orsay, France\\
        F. Bakkali Taheri, G.Doucas, I.Konoplev, A.Reichold, JAI Institute, Oxford, UK\\
		C. Clarke, SLAC, Stanford, USA}

\maketitle

\begin{abstract}
The E-203 collaboration is testing a device on FACET at SLAC to measure the longitudinal profile of electron bunches using Smith-Purcell radiation~\cite{SJPurcell}. At FACET the electron bunches have an energy of 20~GeV and a duration of a few hundred femtoseconds~\cite{intensity}. Smith-Purcell radiation is emitted when a charged particle passes close to the surface of a metallic grating. 
We have studied the stability of the measurement from pulse to pulse and the resolution of the measure depending on the number of gratings used. 
\end{abstract}

\section{Description of the experiment}
A charged particle passing close to the surface of a metallic grating  induces an image charge propagating on the surface of that grating. The couple formed by the electron and its image charge can be seen as an oscillating dipole emitting radiation in all the directions, each one corresponding to a wavelength. If the bunch length is short compared to the emitted wavelength the emission will be coherent. The Coherent Smith-Purcell radiation is related to the longitudinal bunch profile by the following relation:
\begin{equation}
\left(\frac{dI}{d\Omega d\omega}\right)_{N_e}  \approx  \left(\frac{dI}{d\Omega d\omega}\right)_{SP} \left[ N_e + N_e^2 | F(\omega) |^2  G(\sigma_x,\sigma_y) \right]
\end{equation}
Where $| F(\omega) |^2$ is the squared module of the profile's Fourier transform. More details can be found in~\cite{prstab}.

The E-203 experiment installed at FACET (Facility for Advanced Accelerator Experimental Tests) consists in three gratings and a blank piece of metal mounted on a carousel on one side, and eleven pyroelectric detectors on the opposite side, with the beam passing between them. A motorized arm can expose each grating in turn to the beam. In front of the pyroelectric detectors an assembly of filters allows to select the wavelength at which the signal is expected, to increase the signal over noise ratio.
Pyroelectric detectors have been chosen because of their flat response in the far infra-red range where the Smith-Purcell signal is expected in our case (from 20~$\mu$m to 2000~$\mu$m). 
To perform the spectrum measurement we use three gratings with different pitches ( 50~$\mu$m, 250~$\mu$m and 1500~$\mu$m), each one giving access to one part of the spectrum. 
For each of these measurements the background is subtracted by replacing the grating with the blank, and making a measurement with the same filter. 
At the moment, to smooth pulse to pulse variations and to increase the signal over noise ratio we record more than a hundred pulses for each setting of the graying and filters.

The FACET beam line at SLAC (Stanford Linear Accelerator Center) includes a compression chicane so it can deliver to E-203 several different bunch compressions. The FACET facility is described in~\cite{christine}.

\section{Reconstruction of profiles}
\begin{figure}
\begin{center}
\includegraphics[width=6cm]{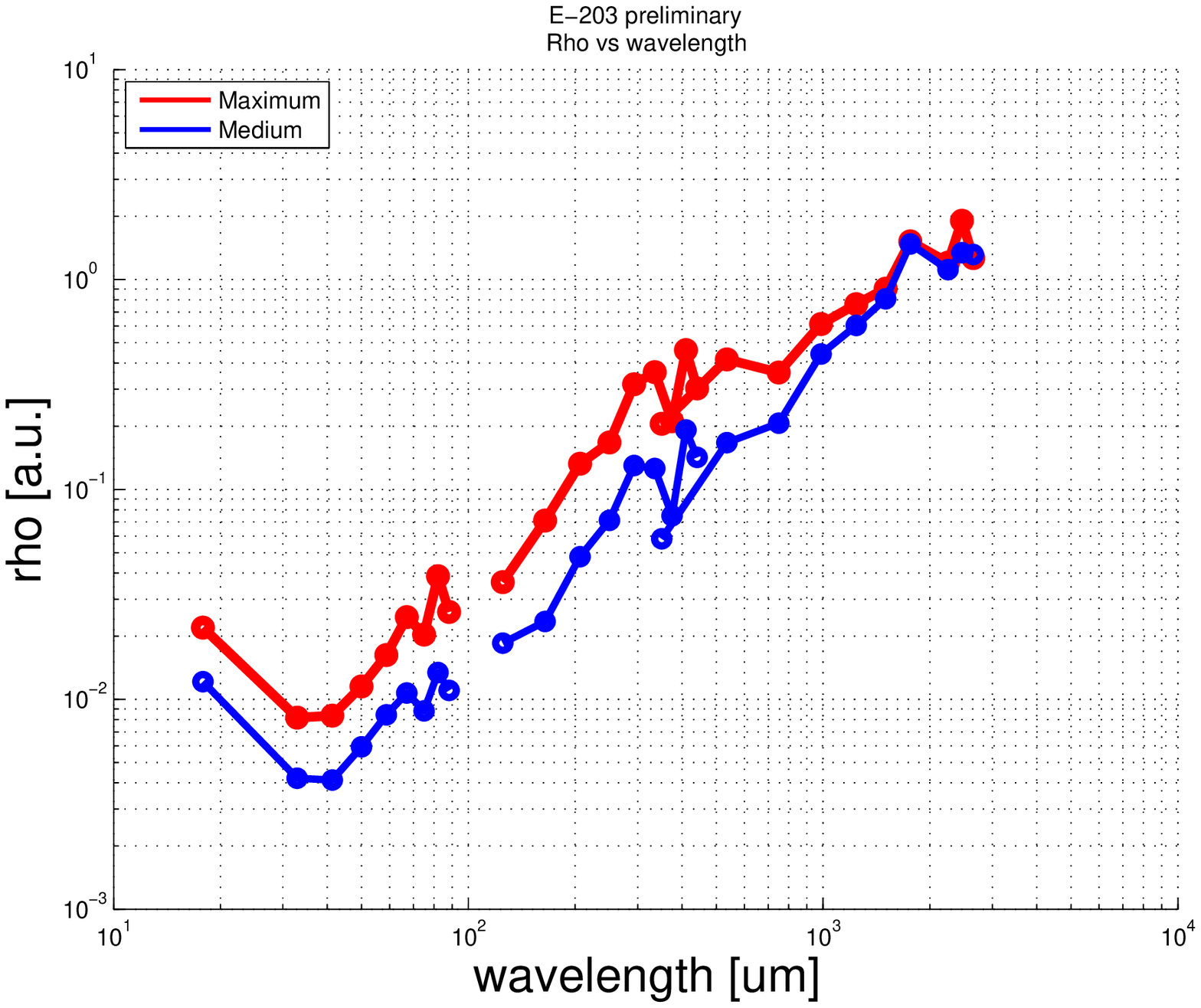}
\includegraphics[width=7cm]{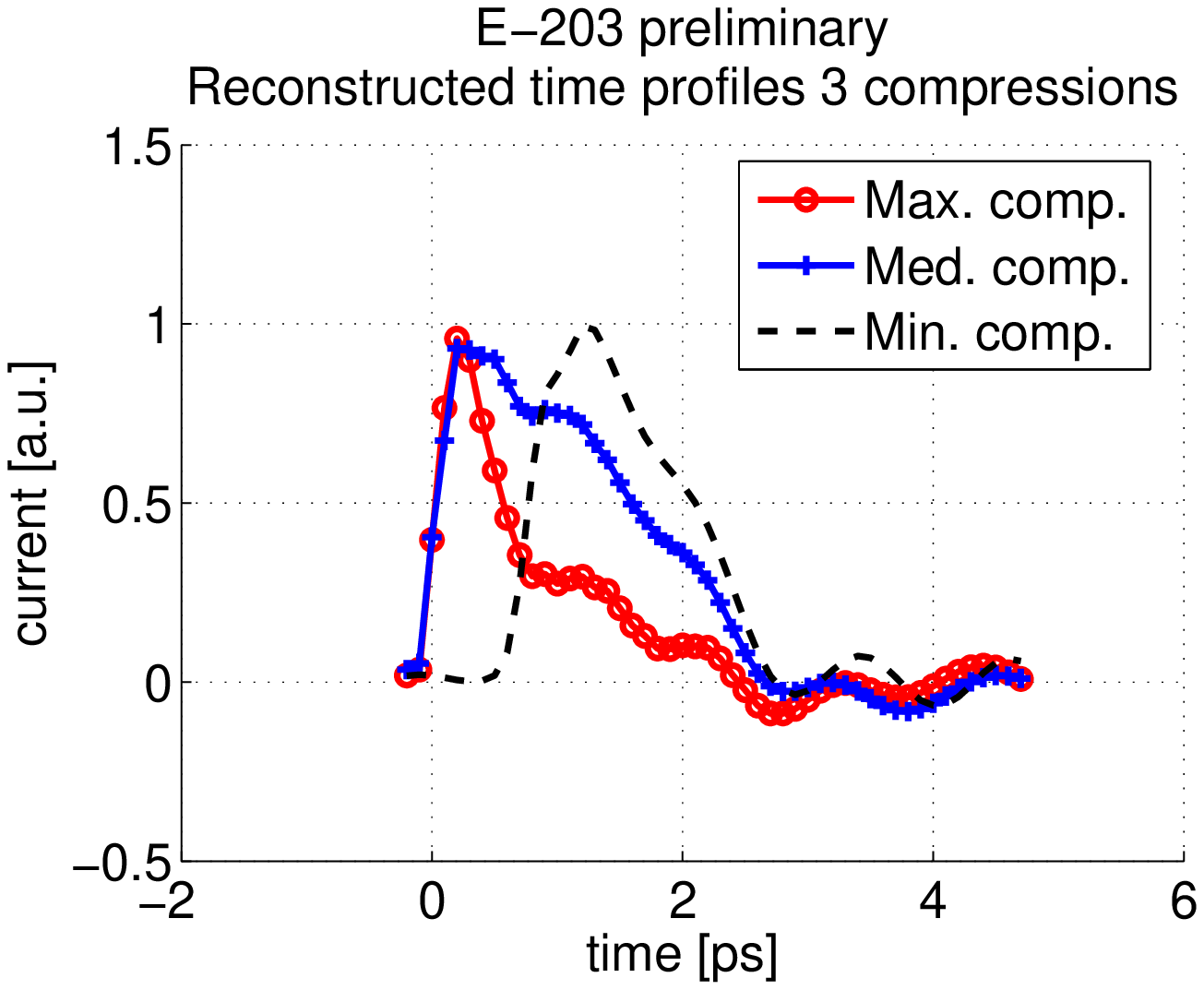}
\caption{Coherent Smith-Purcell radiation spectrum for two different bunch compressions (top) and reconstructed profiles using the minimal phase reconstructed from the Kramers-Kronig relations (bottom). The red and blue lines correspond to two different compression settings, the red being the highest compression. This measurement was done by taking six sets of data of two hundred pulses each. }
\label{3comp}
\end{center}
\end{figure}

To measure the spectrum of the Smith-Purcell radiation emitted by the bunch we perform six sets of eleven data taking using the pyroelectric detectors: three sets for the three gratings with the corresponding filters, and three sets for the background with the blank and the corresponding filters. Then we subtract the background data from the gratings data and we apply corrections to take into account the various efficiencies of the system. 
This gives us the Smith-Purcell radiation spectrum. Examples of such spectrum is shown in figure~\ref{3comp} (top).
 On this image the low wavelengths/high frequencies part (from 10~$\mu$m to 100~$\mu$m) corresponds to the 50~$\mu$m grating, the part from 100~$\mu$m to 500~$\mu$m corresponds to the 250~$\mu$m grating and the high wavelengths/low frequencies part (from 500~$\mu$m to 3000~$\mu$m) corresponds to the 1500~$\mu$m grating.

As shown in equation (1), this spectrum is  the square of the Fourier transform of the bunch longitudinal profile. However to reconstruct it we need to recover its phase. This can be achieved using the Kramers Kronig relations~\cite{desy,ND}. Once this is done, the inverse Fourier transform gives the bunch longitudinal profile.
Examples of such reconstructed profiles are shown in figure~\ref{3comp} (bottom).

 This procedure is described extensively in Victoria Blackmore's thesis~\cite{victoria}.

The validity of this method has been demonstrated in~\cite{prstab}.

\section{Effect of the averaging over many pulses}
Plasma wakefield accelerators have a limited shot to shot stability, therefore there is a strong interest for a single-shot longitudinal bunch profile monitor. Furthermore, such a detector would allow us to do faster measurements at conventional accelerators leading to a more accurate profile, especially at times when the accelerator is slightly unstable.

A required step to do this is to check if a profile reconstructed using six data sets made of a single pulse each is still meaningful and is not dominated by the noise and by accelerator parameters fluctuations.
A simple comparison of the shots with each other is not sufficient to assess the single shot potential, as often several beam parameters change at the same time and may compensate each other.
Therefore, to know if there is a good reproducibility between the shots, we compared directly the final profiles.

\begin{figure}
\includegraphics[width=6cm]{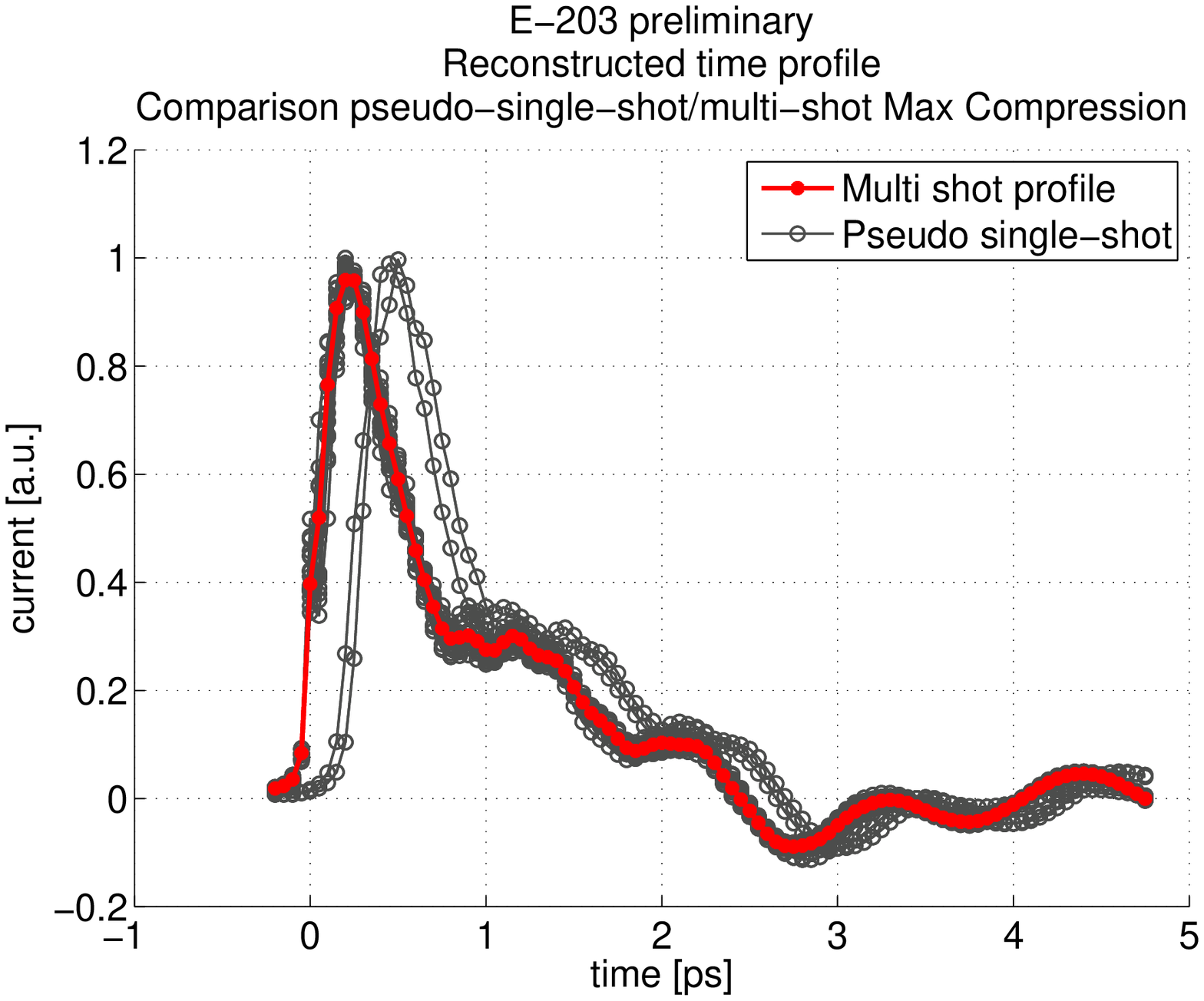}
\includegraphics[width=6cm]{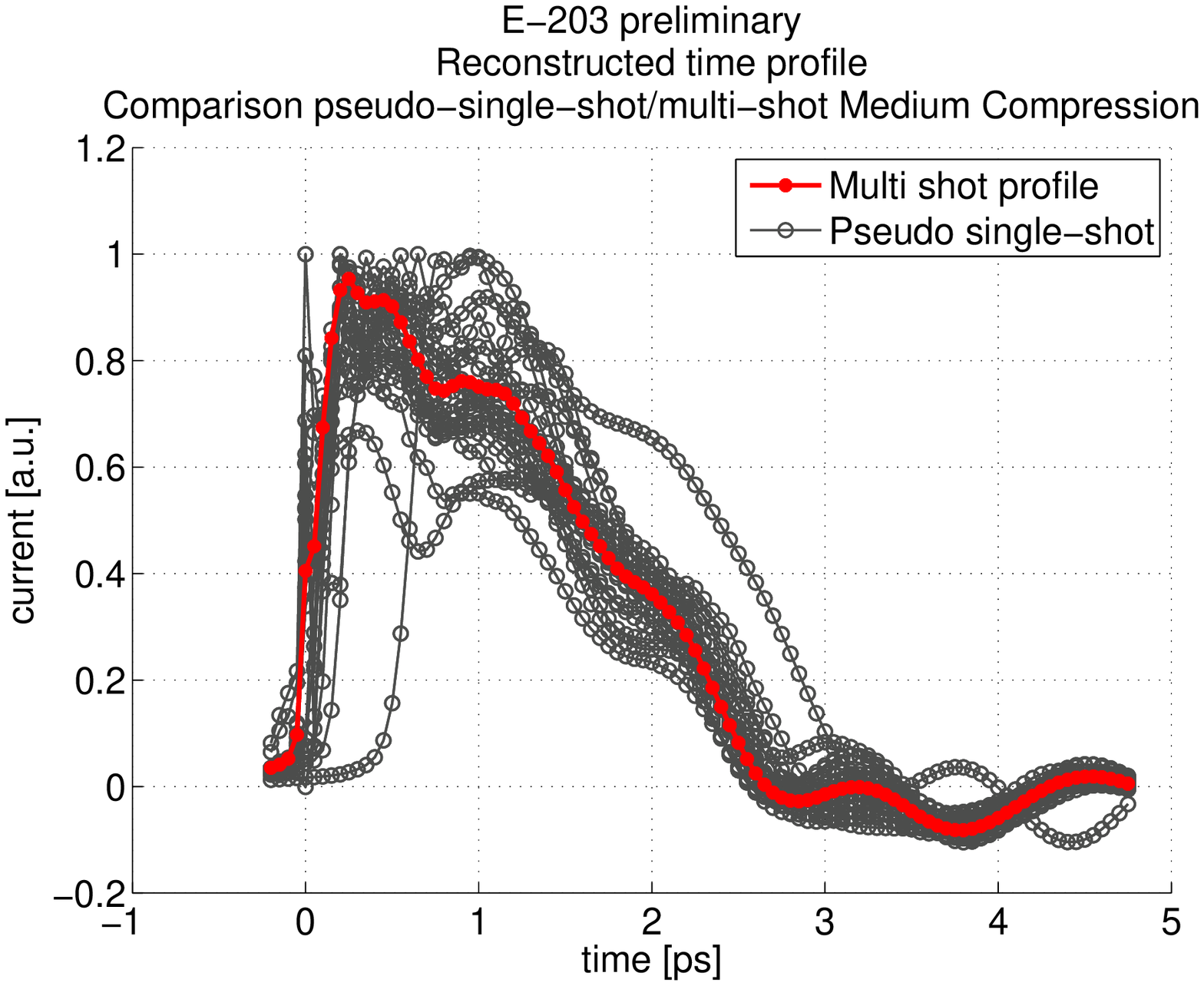}
\caption{Reconstructed profiles for two different bunch compressions (top: highest compression, bottom: medium compression) using an average over 200 shots (red) and the pseudo-single-shot method described in the text (grey).}
\label{repro}
\end{figure}

What we did is that we took one pulse from each data set and used them to reconstruct pseudo-single-shot profiles. A comparison between the profiles obtained by averaging over 200 pulses and by using this pseudo-single-shot method is shown on figure~\ref{repro}. For each compression setting we can see that there is a good agreement between the pseudo-single-shot profiles and the averaged profile.

\section{Use of a single grating}
\begin{figure}
\begin{center}
\includegraphics[width=6cm]{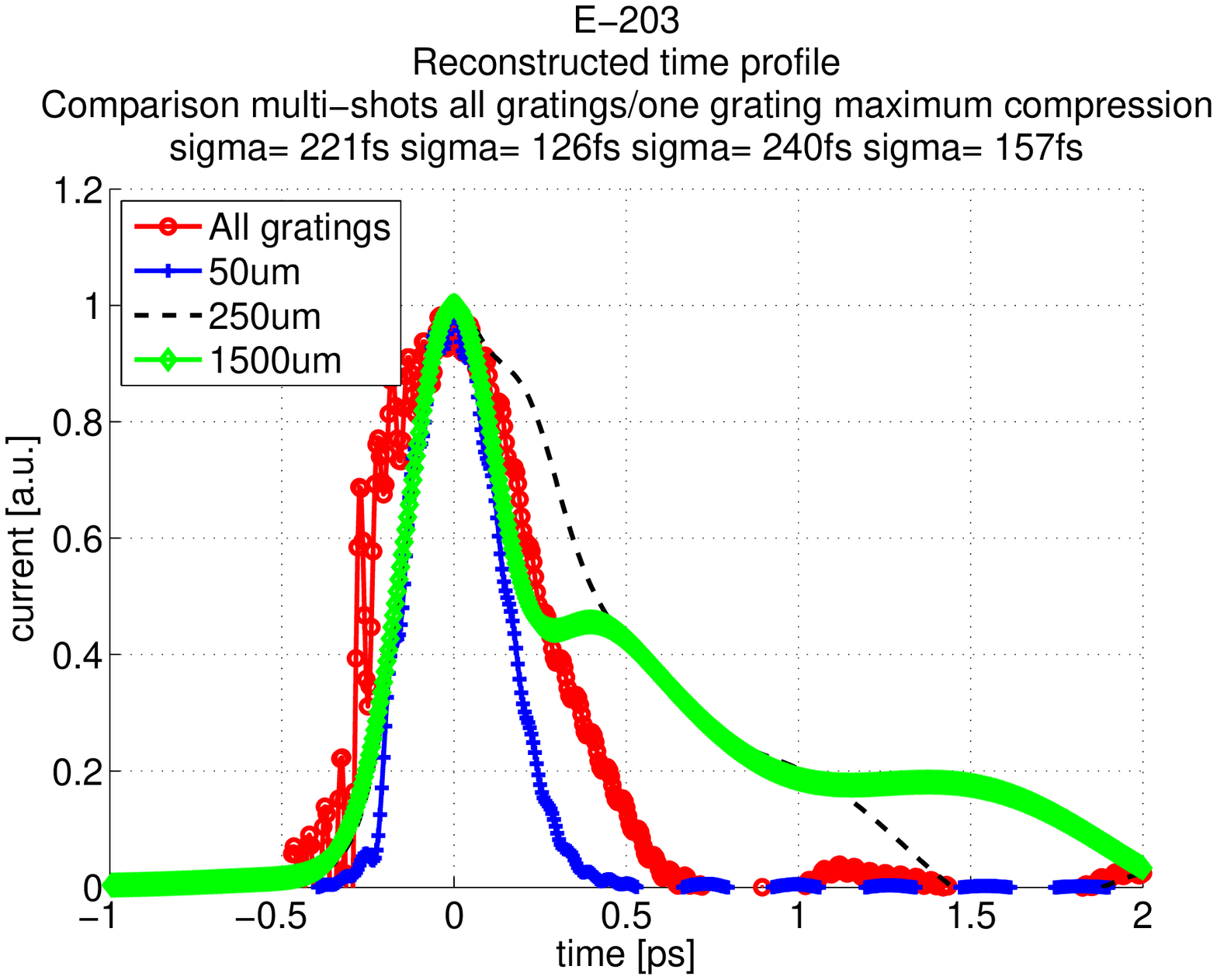}
\includegraphics[width=6cm]{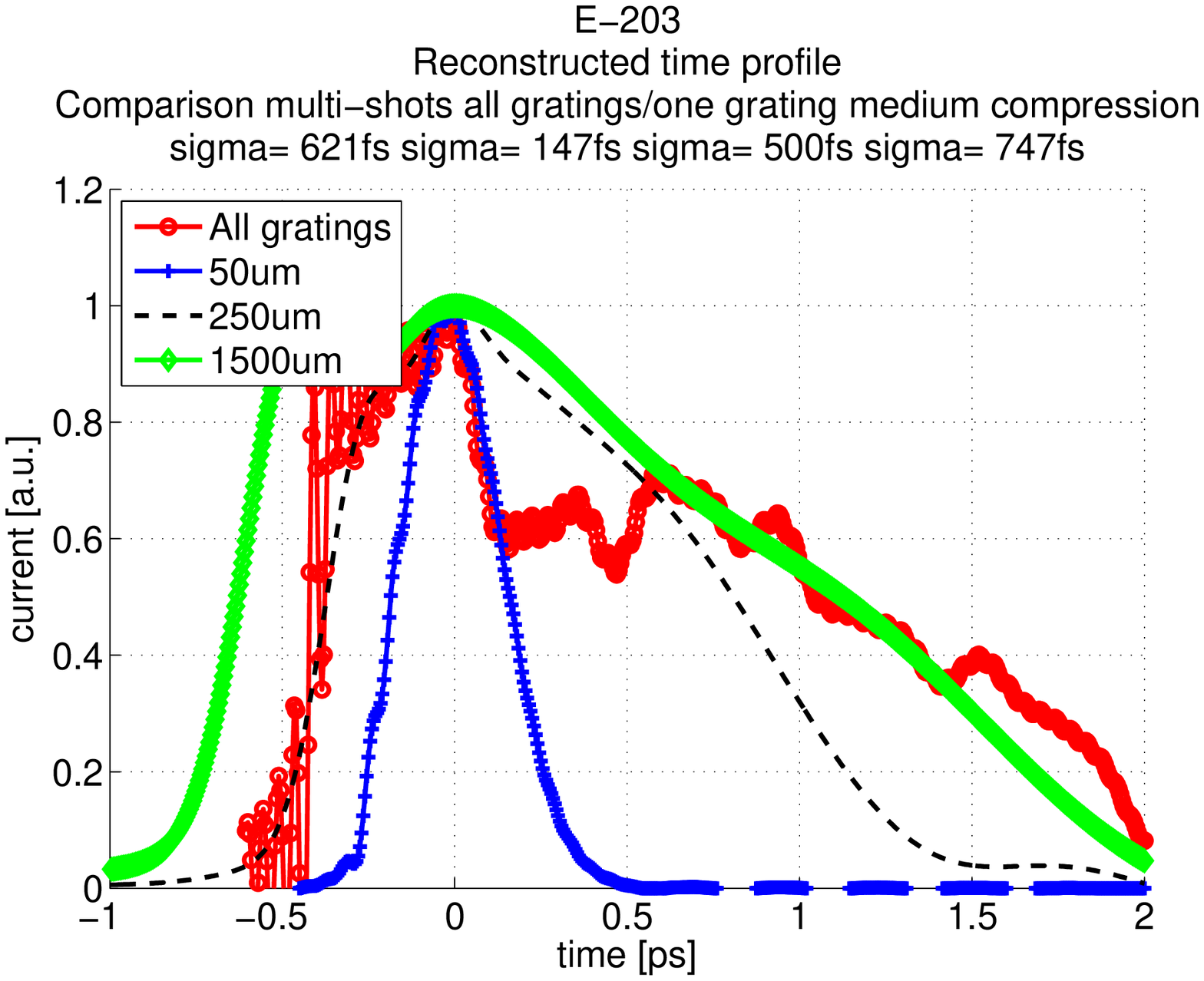}
\caption{Comparison between the profile reconstructed using all the gratings (red) and the profiles reconstructed using only one grating (blue, green and black), for maximum compression (top) and for medium compression (bottom).}
\label{comp_multi_all_one}
\end{center}
\end{figure}

In order to come closer from a single-shot measurement we wonder what would be the effect of using one grating instead of three.
We already know that each grating, with its own pitch, gives access to information with different frequencies however using a single grating reduces the number of data sets needed from six to two. Using these two data sets we reconstructed the profiles corresponding to two different bunch compressions (see figure~\ref{comp_multi_all_one}).
For the highly compressed bunch we can see that all the profiles reconstructed using one grating give the correct peak of the profile however the profiles using the 250~$\mu$m grating and the 1500~$\mu$m grating overestimate the tails. The 50~$\mu$m profile is the closest from the actual profile and this is not surprising because given that the bunch is highly compressed we expect the high frequencies to contribute the most.
For the medium compression of the bunch where the lowest frequencies are expected to dominate, we can see that both the 250~$\mu$m and the 1500~$\mu$m grating profile give good estimates of the actual profile, the 1500~$\mu$m being the closest.

So we can see that provided we have an estimate of the expected bunch length using a single grating can give a good approximation of the bunch profile however the shape will be more accurately reconstructed using several gratings.

\section{Toward a single-shot measurement}
\begin{figure}
\begin{center}
\includegraphics[width=6cm]{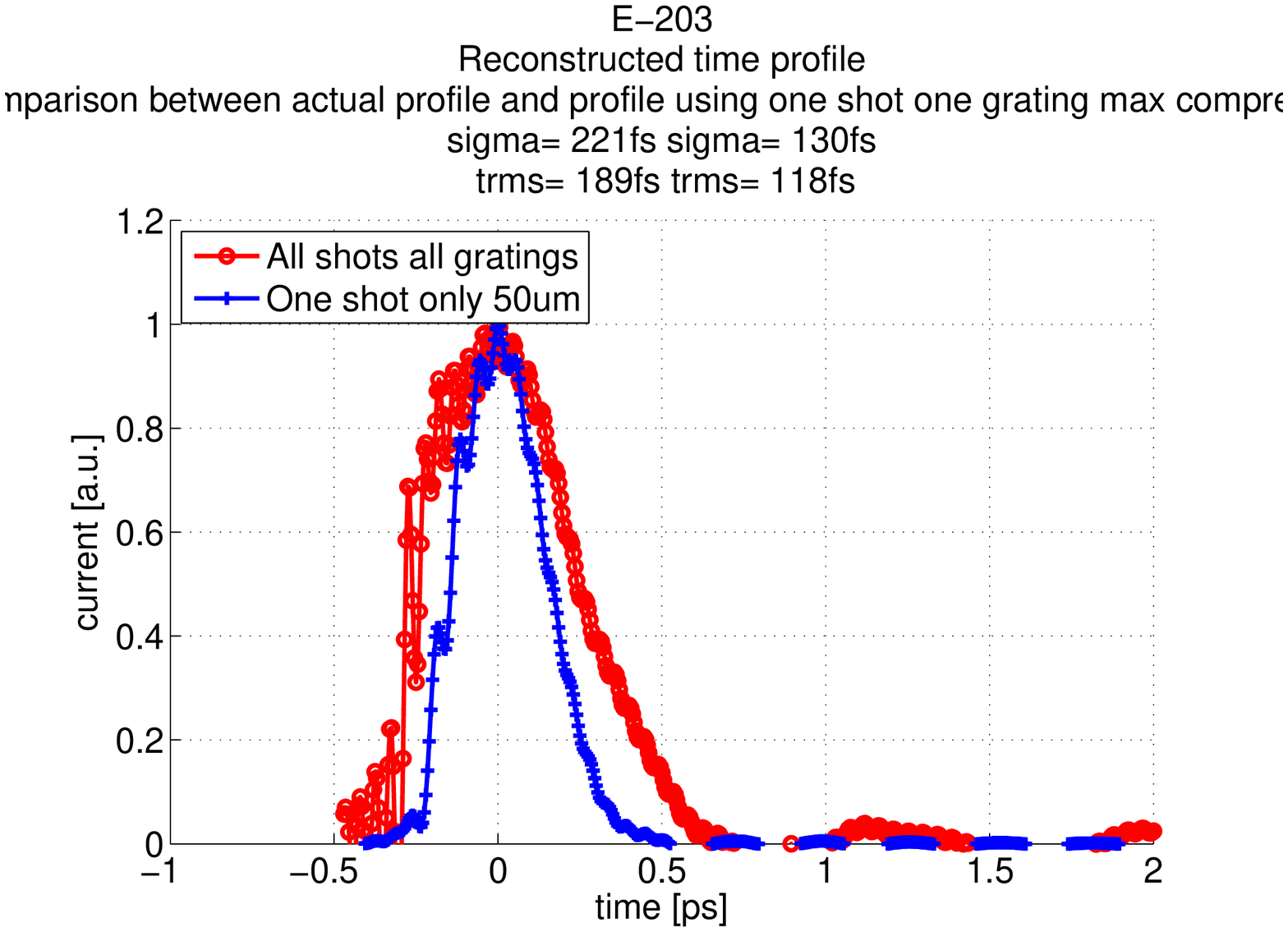}
\includegraphics[width=6cm]{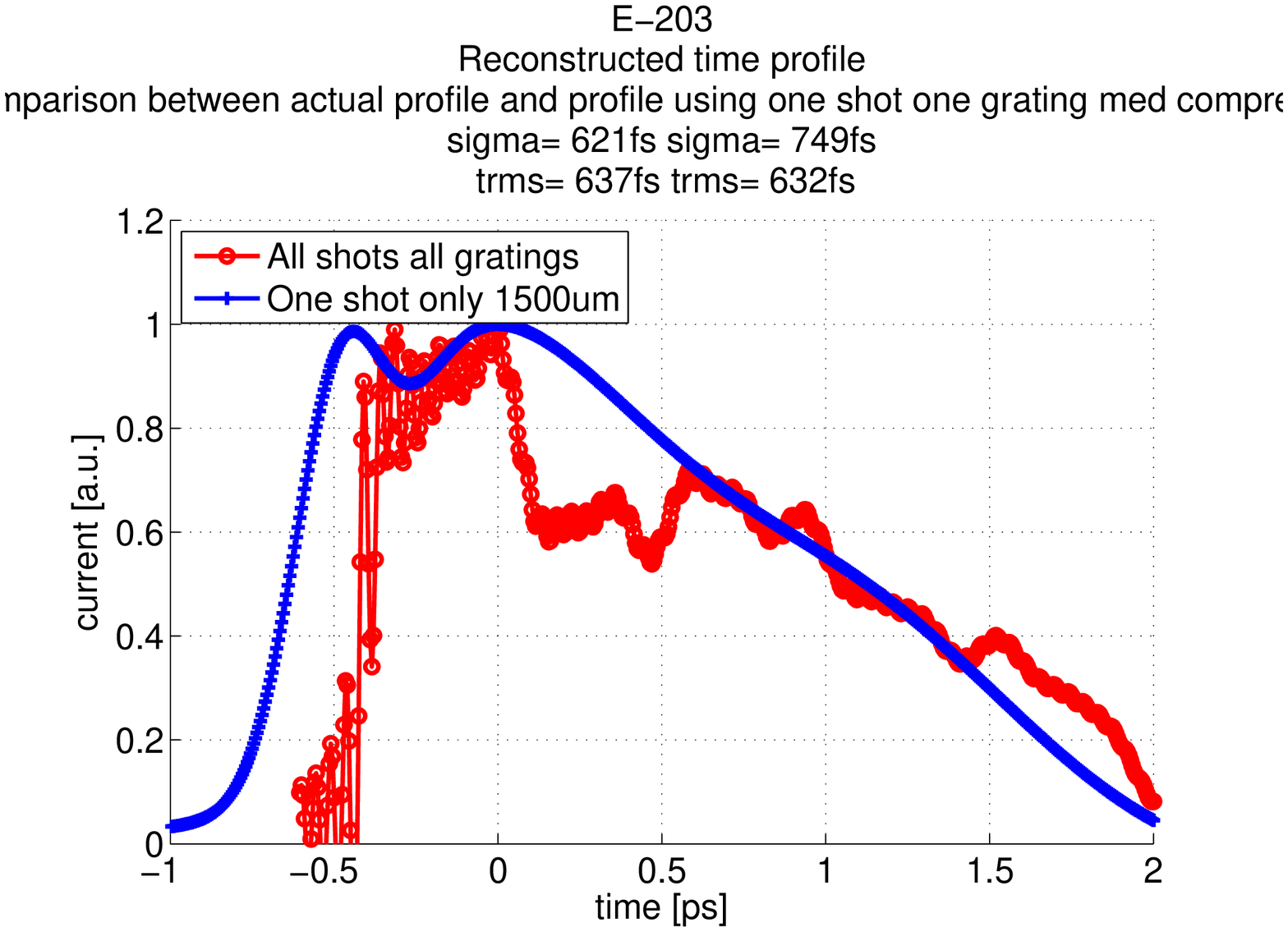}
\caption{Comparison between the actual profile (red) and the profile reconstructed using one grating and one pulse per data set (blue) for maximum compression (top) and medium compression (bottom).}
\label{mix_all_2}
\end{center}
\end{figure}

We have investigated two ways to reduce the number of measurements needed. The first one is to use one pulse per data set instead of more than a hundred, the second one is to use one or two gratings instead of three. Now we will use both these two methods to reduce again the number of pulses needed to reconstruct a profile by taking only one grating and a single pulse each for signal and background. The result of the reconstructions made using this method are shown in figure~\ref{mix_all_2}.
For the medium compression we can see that using one grating (1500~$\mu$m) and one pulse per data set gives a reconstructed profile which is close to the actual one.  We can see the same for the highly compressed bunch using the 50~$\mu$m grating.
By doing so we have reduced the number of pulses used from 1200 to 2.

\section{Conclusion and outlook}
It has been shown that it is possible to use Coherent Smith Purcell radiation to reconstruct a longitudinal bunch profile. But at the moment we use six sets of data (3 gratings and 3 backgrounds set with a blank) and a hundred shots or more per grating to reconstruct this profile. In order to come closer from a single-shot monitor, we have shown that we can reduce the number of pulses from 1200 to 2 while still reconstructing an accurate profile. However to build a real single-shot device we still need to find a way to measure simultaneously the signal and the background. It will require a modification of the geometry of the experiment which may not be possible with our current apparatus at FACET. This is being investigated in the LINAC of the french synchrotron SOLEIL~\cite{last_article}.  A single-shot measurement there is likely to be done in the coming years. We also plan to design a single-shot detector and to test it on a laser driven plasma wakefield accelerator.

\section{acknowledgment}
The authors wish to thank the SLAC FACET team for offering the possibility to perform measurements on their facilities and also for their technical and manpower supports.

\end{document}